\documentclass[twocolumn,aps,prl,superscriptaddress,dvipsnames]{revtex4-1}
\usepackage{graphicx}
\usepackage{color}
\usepackage{amsmath, amssymb}
\usepackage{type1cm}
\usepackage{mathrsfs}
\usepackage{url}
\begin{document}

% Use the \preprint command to place your local institutional report
% number in the upper righthand corner of the title page in preprint mode.
% Multiple \preprint commands are allowed.
% Use the 'preprintnumbers' class option to override journal defaults
% to display numbers if necessary
%\preprint{}

%Title of paper
\title{Prediction of flow dynamics using point processes}

% repeat the \author .. \affiliation  etc. as needed
% \email, \thanks, \homepage, \altaffiliation all apply to the current
% author. Explanatory text should go in the []'s, actual e-mail
% address or url should go in the {}'s for \email and \homepage.
% Please use the appropriate macro foreach each type of information

% \affiliation command applies to all authors since the last
% \affiliation command. The \affiliation command should follow the
% other information
% \affiliation can be followed by \email, \homepage, \thanks as well.
\author{Yoshito Hirata}
\affiliation{Institute of Industrial Science, The University of Tokyo, 4-6-1 Komaba, Meguro-ku, Tokyo 153-8505, Japan}
\author{Thomas Stemler}
\affiliation{School of Mathematics and Statistics, The University of Western Australia, Crawley, WA 6009, Australia}
\affiliation{Potsdam Institute of Climate Impact Research, P.O. Box 60 12 03, 14412 Potsdam, Germany}
\author{Deniz Eroglu}
\affiliation{Potsdam Institute for Climate Impact Research, P.O. Box 60 12 03, 14412 Potsdam, Germany}
\affiliation{Department of Physics, Humboldt University of Berlin, Newtonstrasse 15, 12489 Berlin, Germany}
\author{Norbert Marwan}
\affiliation{Potsdam Institute of Climate Impact Research, P.O. Box 60 12 03, 14412 Potsdam, Germany}
%\email[]{Your e-mail address}
%\homepage[]{Your web page}
%\thanks{}
%\altaffiliation{}

%Collaboration name if desired (requires use of superscriptaddress
%option in \documentclass). \noaffiliation is required (may also be
%used with the \author command).
%\collaboration can be followed by \email, \homepage, \thanks as well.
%\collaboration{}
%\noaffiliation

\date{\today}

\begin{abstract}
Describing a time series parsimoniously is the first step to study the underlying dynamics. For a time-discrete system, a generating partition provides a compact description such that a time series and a symbolic sequence are one-to-one. But, for a time-continuous system, such a compact description does not have a solid basis. Here, we propose to describe a time-continuous time series using a local cross section and the times when the orbit crosses the local cross section. We show that if such a series of crossing times and some past observations are given, we can predict the system's dynamics with fine accuracy. This reconstructability does not depend strongly on the size nor the placement of the local cross section if we have a sufficiently long database. We demonstrate the proposed method using the Lorenz model as well as the actual measurement of wind speed.
\end{abstract}

%\pacs{05.45.Tp,87.19.ls,87.19.lo,92.60.Gn}
% insert suggested keywords - APS authors don't need to do this
%\keywords{}

%\maketitle must follow title, authors, abstract, \pacs, and \keywords
\maketitle

{\bf Current developments of measurement techniques and hardware enable us to record time-continuous data with very high sampling rates for long times. To understand such data intuitively, we need to describe the data parsimoniously so that such description can reproduce the original time series. Thus, here we propose to represent such time-continuous data only by the series of times when the orbit passes a local cross section. We show that a time series together with some measurements of the past dynamics of the system is sufficient information to predict the future dynamics of the system. There are two applications: an immediate application is to save, or send, time-continuous data using small memory or low channel capacities and therefore make big data more manageable; a more abstract application is to approximate details in a time series when only few observations are possible, based on detailed measurements taken during a different time.}

%General Motivation
Recent advances in measurement technology enable us to record data of  time-continuous systems with very high sampling rates. While these advances are welcome, there are several problems in accessing, analysing and even storing such data sets. These problems are nowadays summarised under the umbrella of \emph{big data} and arise in diverse areas ranging from bioinformatics to systems' health engineering~\cite{Bio,health}.
%\todo[inline]{It would be good to add here one or two examples of big data problems where our method might be useful.}
Here we are presenting a method based on the recurrence of dynamical systems to categorise and rank such data sets. Ranking the recurrence times of our current state against the recurrence times of the big data set, allows us to predict the future dynamics of the system.

%Our method in simple words (next 2 paragraphs)
Our research is motivated by recent achievements in the time series analysis of non-uniformly sampled data~\cite{Irregular1, Irregular2, hirata2009, hirata2012, hirata2014, ozken2015, eroglu2016}. We assume, we have a time-continuous dynamical system, that we are able to measure for a certain amount of time. This observational series of the high dimensional state space is our database. After recording the full dynamics in the phase space for our database, we no longer measure the full dynamics but record the crossing times. These crossing times are the times when the trajectory of the system passes through a local cross section on the attractor of the dynamical system. This data set is our non-uniform sampled data set, which we use to estimate the current state of the system. This estimation is done by finding similarities between the current crossing time sequence and the crossing times calculated from the database of the full dynamics. 

%This distance provides a metric that can be used to reveal the information in the crossing time series. 
To detect the similarity we facilitate the Victor and Purpura distance~\cite{victor1997}. The information contained in such crossing times has been studied since Poincar\'e~\cite{him} and is the foundation of recurrence--based time series analysis~\cite{NM}. The Victor and Purpura distance~\cite{victor1997} offers a natural metric to rank and detect similarities between consecutive time windows of the crossing times~\cite{victor1997,hirata2009,hirata2012,hirata2014,ozken2015,hirata2016,eroglu2016}. Using cross prediction~\cite{levanquyen1999}, we exploit these similarities to estimate the current state and predict the future system's dynamics.  

%Give a plan and overview
Our paper is organised as follows. After giving a short summary of our method, we first give details on the information contained in the crossing time series and introduce the Victor and Purpora distance. Then we introduce the cross prediction method and illustrate our method using the low-dimensional Lorenz system~\cite{lorenz1963}. In addition we apply our method to predict actual wind speed data. %At last, we also discuss a possibility for using the proposed method to overcome gaps within observations such as in satellites images partly covered by clouds~\cite{liu2017}, the historical series of sunspot numbers~\cite{clette2014} or historical phenological data such as the start of the cherry blossom in Japan~\cite{aono2008}.

%The prediction and the role of the database
Predicting the dynamics of the system will be done by facilitating the information in the database. This record of the past dynamics is used in two ways. First to compare the crossing time series of the database with our current crossing time series. Second we predict the future dynamics by approximating the future state of the system from the past dynamics in the database. The length of the database is critical and determines the quality of our prediction. The recording time has to be chosen long enough to allow first a good state estimation and then a realistic approximation of the future dynamics. Since our state estimation as well as the cross prediction rely on similarity between past and future dynamics the longer the record the more reliable the prediction becomes. One of the main results of this paper is that even for realistic short recording times our method results in quite reliable predictions. 

%Defining the database and recording time as well as some jargon
Formally we assume a dynamical system $f_t:M\rightarrow M$, where $M$ is an $m$-dimensional manifold and $f_t$ is the diffeomorphism representing how a point $y$ in $M$ moves to a point $f_t(y)$ in $M$ after some time $t$. Our database contains the full dynamics of $y(t)$ for $t=[0,\Delta t, 2\Delta t, \ldots{}, L]$, where $L$ denotes the recording time and $\Delta t$ is the sampling time. If the full dynamics cannot be directly measured, one has to reconstruct the dynamics using delay embedding~\cite{takens1981,sauer1991}. We will discuss this further, when we introduce our environmental example of predicting wind speeds. 

\begin{figure}[tb]
\includegraphics[width=\linewidth]{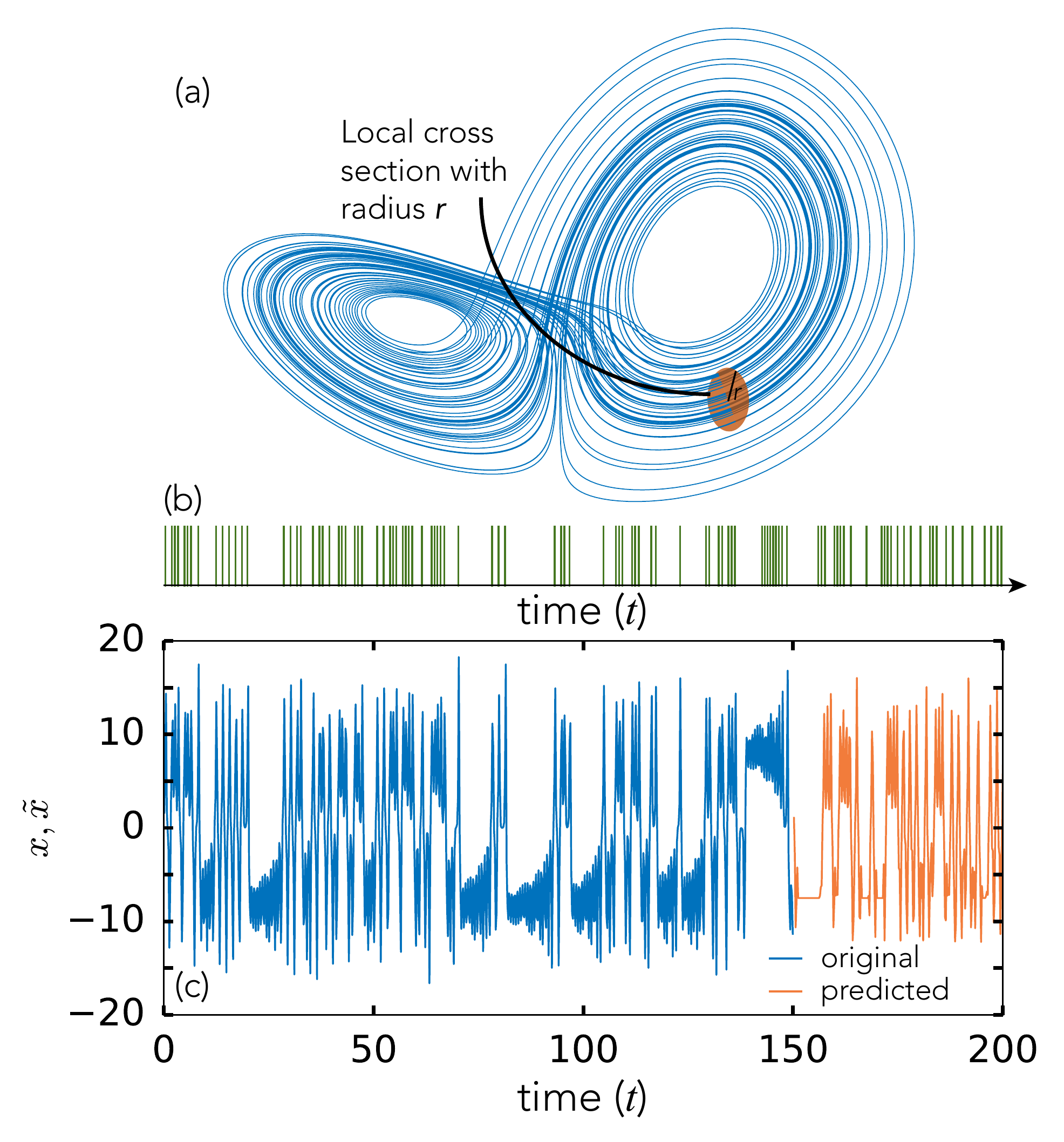}
\caption{\label{figure1} A schematic overview of our prediction method. (a) Local section $\Sigma$ with radius $r$, oriented perpendicular to the flow; (b) spike train having a non-zero amplitude every time the trajectory passes through $\Sigma$; (c) $x$-component of the Lorenz system (blue) and $\tilde{x}$ our prediction (green). The blue part of the trajectory represents the database used for the cross prediction.}
%\todo[inline]{maybe we cold also plot the predicted states in the phase space (Fig. 1a)?}
\end{figure}

%Crossing time series and local section
From $t>L$ onwards we do not have access to the full dynamics anymore, but measure the times $\{t_i\}$ the trajectory $y(t)$ crosses a local section $\Sigma\subset M$, e.g. $y(t_i)\in\Sigma$. We choose $\Sigma$ to be perpendicular to the flow \textcolor{black}{at a point of the trajectory} and the size of $\Sigma$ is given by its radius $r$ (c.f. Fig.~\ref{figure1}(a)). \textcolor{black}{(In practice, we choose a point in the time series and calculate the direction the point is moving. This direction gives the the normal vector for our local cross section $\Sigma$.)} A lot is known about such crossing time series. For example it can be shown that a $d$-dimensional vector $I_i = (t_{i+1}-t_{i},t_{i+2}-t_{i+1},\ldots,t_{i+d}-t_{i+d-1})$ provides a one--to--one embedding of the dynamics under mild conditions relating to periodic points in the phase space~\cite{sauer1994,huke2007,smale2011} as long as $d>2m$. We exploit this result but instead of measuring $d+1$ consecutive crossings, our measuring time $w$ is chosen so that our crossing time series always satisfies the inequality $d>2m$:
\begin{equation}
\label{equation1}
\max_{i} \{t_{i+2m+1}-t_i\} < w.
\end{equation}
%Estimate based on database
We therefore record the crossing time series of our system for $w$ time units. To determine $w$ we use the time series $y(t)$ in the database and measure the crossing time series $\{t_i\}_{DB}$. This set of times $\{t_i\}_{DB}$ is going to be used to determine $w$. Therefore $w$ is the maximum time span needed in the recorded history (the database) to satisfy the condition $d>2m$. Note that, especially if the recording time $L$ is short, choices of $w$, as the upper bound of eq.~(\ref{equation1}), should be taken conservatively. As we will see, in practice one can even use a small value for $w$, e.g. an upper bound for only a section $i$ of the time series, and still get a good prediction. We discuss techniques used to improve the prediction for such choices of $w$ later. 

%State estimation and intro to VP distance
Given the current crossing time series $\{t_i\}$, we use the Victor and Purpura distance metric~\cite{victor1997} to find similar sequences of length $w$ in $\{t_i\}_{DB}$. The closest sequences in $\{t_i\}_{DB}$ provide the base for our state estimation and the prediction of the dynamics for $t>L$. To apply the Victor and Pupura distance we represent the crossing time series $\{t_i\}$ and $\{t_i\}_{DB}$ as their corresponding spike trains. The value of such spike trains are everywhere 0 and only equal to 1 when the trajectory $y(t)\in\Sigma$ (c.f. Fig.~\ref{figure1} (b)). The Victor and Purpura metric determines the minimum cost to convert one spike train into another one. This metric has been used extensively in the context of neuroscience, e.g.~\cite{hirata2009}, where the spike trains naturally arise as the output of idealised neurons. 

%Details of VP next three (!) paragraphs, but we need it
Let the current spike train be $U = \{u_\xi| \xi=1,2,\ldots,\Xi\}$ and let $V = \{v_\theta| \theta=1,2,\ldots,\Theta\}$ represent one of the possible spike trains of the database. $u_\xi$ denotes the time for the $\xi$th event within $U$ and $v_\theta$ is the time for the $\theta$th event within $V$. Thus, given two initial conditions $x_u,x_v \in X\subseteq M$ at the beginnings of the two windows, we have $f_{u_{\xi}}(x_u),f_{v_{\theta}}(x_v) \in \Sigma$ for each $\xi$ and $\theta$. When we apply the shift \textcolor{black}{of events}, we pair up one event $u_\xi$ from $U$ with another $v_\theta$ from $V$. Therefore each $u_\xi$ and $v_\theta$ may not belong to more than one pair. In addition, we define $C$ as the set of such pairs taken from $U$ and $V$.  With this we can define the Victor Purpura distance as:
\begin{equation}
\label{equation2}
\delta(U,V)=\min_{C} \{\sum_{(u_\xi, v_\theta) \in C} \lambda |u_\xi-v_\theta|+\Xi+\Theta-2|C|\}.
\end{equation}
This edit distance $\delta$ satisfies the necessary three conditions for a metric; (i) non-negative or ,zero iff two time windows are identical, (ii) symmetric, and (iii) the triangle inequality.

Intuitively the metric can be understood by taking into account the possible edits that can transform $U$ into $V$: either we can align event $u_\xi$ and $v_\theta$ or we have to delete / create events in the spike trains. Both these edits appear in the RHS of (\ref{equation2}). The cost of aligning two events is proportional to the time difference between the two events, e.g. $\lambda| u_\xi-v_\theta|$ controls the cost. Since the cost of creating and deleting events is chosen to be equal to 1, $\delta$ depends on the difference between the total number of events -- given by $\Xi+\Theta$ -- and twice the number of possible pairs -- $2|C|$. 
%\todo[inline]{This preceding sentence is difficult to understand.}
\textcolor{black}{This Victor Purpura distance has been used to evaluate synchonization among neurons~\cite{macleod1998,nomura2015}.}
Further details and a thorough mathematical description of the Victor Purpura distance can be found in the literature~\cite{suzuki2010,ozken2015,hirata2015}.

%What we found: state estimation in reality
Comparing the current crossing time sequence in this way with all the crossing time sequences in the database, we identify the 10 sequences that are closest in terms of their distance $\delta$. The average of their end points is going to be the state estimate at the current time. For extremely long database recording times and very high data precision one might use just the end point of the sequence with minimal $\delta$ as the current state estimate. Under realistic conditions, however, using the average \textcolor{black}{and standard deviation} of about 10 sequences with similar recurrence properties \textcolor{black}{-- in the Victor Purpura distance sense -- improves} the stability of the prediction \textcolor{black}{and helps us to evaluate its reliability}. \textcolor{black}{One can understand the Victor-Purpura distance for a spike train as being similar to the Euclidean distance for usual vector spaces.}

%Forecasting (I find my way of writing this stuff better to understand )
Using cross prediction has the advantage that we do not require access to the equations governing the dynamics of $y(t)$. Instead the prediction of the future dynamics is provided as the average of the recorded evolution of the 10 states defining our state estimate. We define $t_0$ as the current point in time and use $h=1,2,\ldots{},10$ as the index of the 10 states. Consequently \textcolor{black}{$t_{0,h}$} denotes the end points of the 10 most similar sequences in the database. Then the estimation of the current state is:
\begin{equation}
\label{equation3}
\textcolor{black}{\tilde{y}(t_0|t_0)}=\frac{1}{10}\sum_{h=1}^{10} y_h(t_{0,h}).
\end{equation}
Similarly the prediction \textcolor{black}{of $\tilde y$ for} $n\Delta t$ time steps in the future is:
\begin{equation}
\label{equation4}
\textcolor{black}{\tilde{y}(t_0+n\Delta t|t_0)}=\frac{1}{10}\sum_{h=1}^{10} y_h(t_{0,h}+n\Delta t),
\end{equation}
and its corresponding prediction uncertainty can be measured by the variations of the 10 sequences:
\begin{equation*}
%\label{equation5}
\textcolor{black}{\sigma_{\tilde{y}}(t_0+n\Delta t|t_0)}=\sqrt{\frac{1}{10}\sum_{h=1}^{10} (y_h(t_{0,h}+n\Delta t)-\tilde{y}(t_0+n\Delta t))^2}.
\end{equation*}
We use the uncertainty of our prediction to accept or reject the prediction of $t_0+n\Delta t$. The main issue we will have is that the ensemble of the 10 sequences with the lowest $\delta$ might lose coherence, because they might be close in the Victor-Purpura sense, but not all of them have to be close to the system's state on $\Sigma$. When this happens the uncertainty of the prediction grows: \textcolor{black}{$\sigma_{\tilde{y}}(t_0+n\Delta t|t_0)>\sigma_{\tilde{y}}(t_0+n\Delta t|t_0+n\Delta t)$}. In such a case we would reject the prediction and instead let \textcolor{black}{$\tilde{y}(t_0+n\Delta t|t_0+n\Delta t)$ be our prediction and replace $t_0$ by $t_0+n\Delta t$ to reinitialise for further prediction.}
% until the uncertainty of our prediction becomes less then the uncertainty that led to the rejection. To prevent divergence of the whole ensemble we compare the uncertainty of the 10 ensemble members against the future prediction $\tilde{y}(t)$ for $t=[t_0, t_0+w/2]$. In our experiments $w/2$ is normally short and about one to five periods of the system's characteristic oscillation. Since $w$ is also the length of our spike trains, we reduce the number of free parameters in our algorithm. If a member's uncertainty exceeds a certain threshold within this period, we will replace this ensemble member with the member that has the next lowest $\delta$ and is not already part of the ensemble.

%As a final step, we need to choose a time $\Delta T$, after which we are going to reinitialise our prediction. Since our spike trains have a certain length, a reinitialisation shortens and lengthens the gaps at the beginning and end of the sequence. Since alignment of these gaps and spikes determines the Victor-Purpura metric used to estimate the state, $\Delta T$ has direct impact on the prediction quality. The values for $\Delta T$ we used in our numerical experiments below is small compared to the average period of the system. 
%{\red
%At each increment $\delta t$, we evaluate whether 
%}

Especially when $L$ is short, it can be necessary to work with a small value of $w$. For such $w$ the inequality (\ref{equation1}) will not be satisfied for all times, but only for some part of the time series in the database. To still achieve a good state estimation, we consider that just before or after the current time window, there is or is not a spike. For our current spike train, we therefore have the option that a spike does (not) appear just before (after) the beginning (end) of the window. We combine these 4 options with the corresponding 4 options of our database $\{t_i\}_{DB}$ and determine the $\delta$ of the 16 possibilities. As we see below, including these 16 possibilities in our $\delta$ minimisation greatly improves the accuracy of the prediction.
\begin{figure}[tb]
\includegraphics[width=\linewidth]{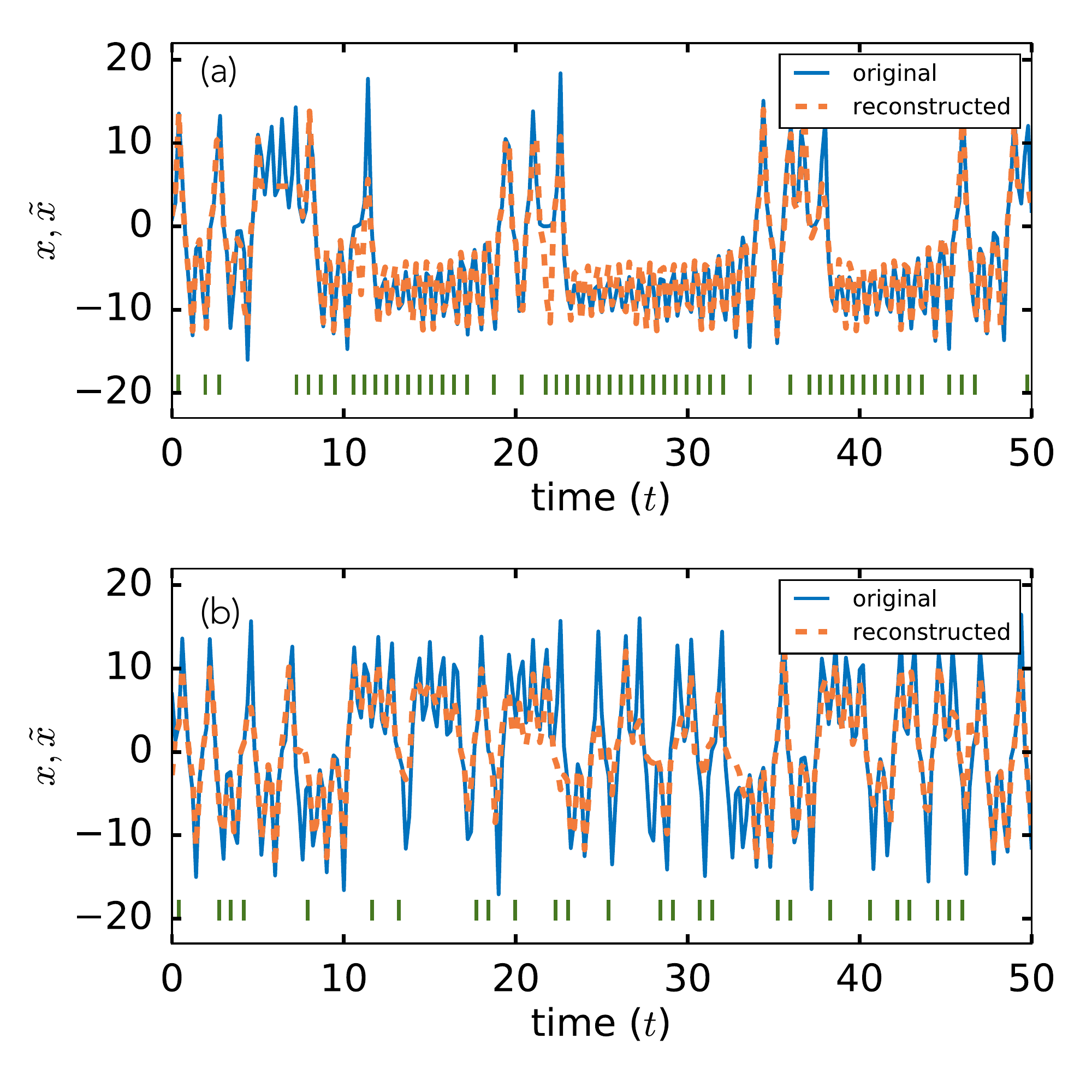}
\caption{\label{figure2}Comparison between the predicted $\tilde{x}$ (dashed line) and the original $x$ (solid line) time series of the Lorenz system: (a) recording time of the database $L=150$, radius of local section $r=10$, upper bound of eq.~(\ref{equation1}) $w=2$ and \textcolor{black}{$\Delta t=0.2$}; (b) $L=4950$, $r=3$, $w=10$ and \textcolor{black}{$\Delta t=0.2$}; at the bottom of both panels the corresponding spike trains are shown and $\tilde{x}(t_0)$ is given by eqs.~(\ref{equation3},\ref{equation4}). }
\end{figure}

Such a prediction can be seen as the orange line in Fig.~\ref{figure1}(c). Clearly visible are the time periods during which we rejected the prediction (see the instances of constant amplitude). The coincide with periods of low or no recurrence, that are large gaps in the train sequence in panel (b). For practical applications we would use a larger value of $w$ to eliminate these areas of prediction rejection. To qualitatively assess the prediction we evaluate the cross-correlation coefficient $R(y,\tilde{y})$ between the prediction and the true dynamics of the system. For each $t_0$ we are going to predict the dynamics using eq.~(\ref{equation4}) and evaluate $R(y,\tilde{y})$. 

%Explain what we do
As a first application of this method, we apply our algorithm to the Lorenz 1963 system~\cite{lorenz1963}, using the standard parameters $\sigma=28$, $\rho=10$ and $b=8/3$. We are going to systematically vary the radius $r$ of the local section $\Sigma$ and the database recording time $L$. These parameter changes directly impact on the upper bound $w$ of eq.~(\ref{equation1}). \textcolor{black}{We calculated statistics $Cv$ and $Lv$ from these spike trains based on Ref.~\cite{shinomoto2003}, which evaluate the global and local variabilities for spike trains for judging whether the spike trains follow a Poisson process or not. If a spike train follows a Possion process, its $Cv$ and $Lv$ fluctuate around 1~\cite{shinomoto2003}. We found that $Cv$ and $Lv$ were $0.8603\pm 0.1904$ and $0.6276\pm 0.1790$, respectively, over 10 samples. While the estimate for $Cv$ was not significantly different from 1, that for $Lv$ was significantly different from 1 ($P \approx 0.019$), meaning that the spike trains are likely to be different from a Poisson process, or a standard point process~\cite{shinomoto2003}.}

\begin{figure}[tb]
\includegraphics[width=\linewidth]{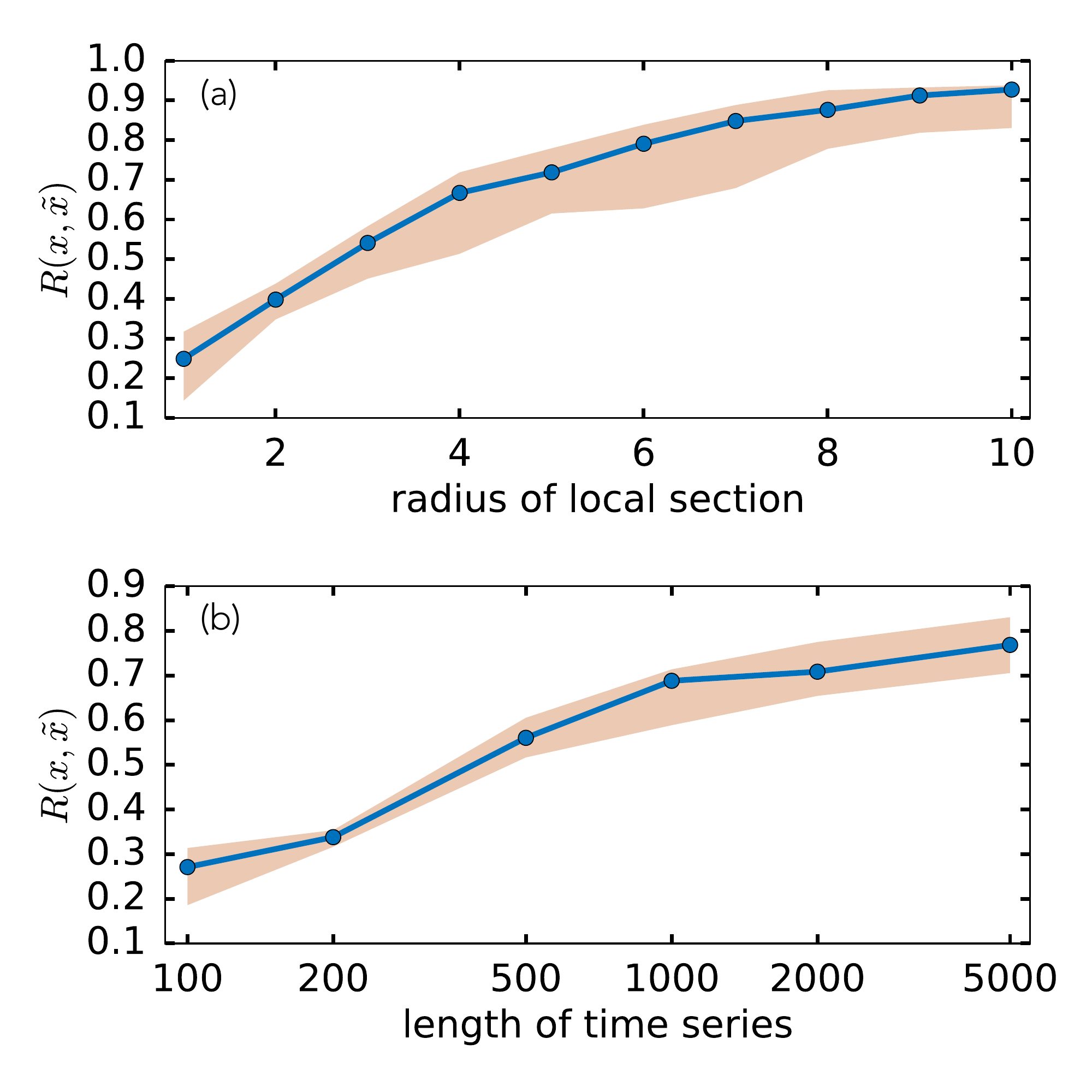}
\caption{\label{figure3}The cross-correlation coefficient $R(x,\tilde{x})$ between the prediction and the true dynamics in dependence of the size of the partition and recording length. Cross-correlation was calculated for a prediction of 50 time units. Shown is the average correlation for 10 randomly placed $\Sigma$s as well as the 50\% confidence interval. (a) fixed upper bound $w=2$ and $L=150$ while varying the size of the portion and (b) $w=10$ and a fixed radius $r=3$ of the partition while varying the recording length $L$. \textcolor{black}{N.B. that we tend to have positive correlation coefficients because we predicted the values generated from the true dynamics based on spike trains.}}
\end{figure}
%Truthfulness of estimate equation 3 Figure 2
In Fig.~\ref{figure2} we see two outputs of our algorithm in comparison with the true $x$-component of the Lorenz system. Since it is hard to visualise the state estimates together with their predictions, we instead only show the state estimates together with the prediction up to the next estimate. As a reinitialisation time we used $\Delta t=0.2$ in both experiments. Since this value is very small, the prediction is never rejected and the data shown highlights the accuracy of our state estimate more than the quality of our prediction. The main difference between the two data sets shown in panel (a) and (b) is the recording time $L$ and the radius $r$ of the local section $\Sigma$. We use $r=10$ in panel (a) and $r=3$ in (b). Consequently the recurrence rate with which the trajectory returns to $\Sigma$ is much higher in (a) than in (b) as we can see from the spike trains below the trajectory. Given the high recurrence rate we need a short recording time $L=150$ for $r=10$, while for $r=3$ we need longer recording times ($L=4950$). Similarly the recurrence rate helps us to chose the upper bound $w$ of eq.~(\ref{equation1}): (a) $w=2$ and (b) $w=10$. The comparison of the true and predicted dynamics clearly shows that our algorithm can be optimised to work for both choices of $r$ and approximates the dynamics of the system well. 

The large deviations of $\tilde{x}$ in Fig.~\ref{figure2} (a) around $t=12$ and $22$ are caused by our choice of $w=2$ and a very short recording time $L=150$. Given $w=2$, it is not always possible to satisfy the condition (\ref{equation1}). For that reason our prediction diverges from the true dynamics. In addition the short recording of the past dynamics in combination with the cross-prediction technique does not allow us to get the amplitude of some of the oscillations right, e.g. $25<t<30$. As we can see for $L=4950$ and $w=10$ these problems disappear and we get better estimates and predictions (Fig.~\ref{figure2} (b)). While the recurrence rate is lower, we have more knowledge about the past dynamics and the higher value of $w$ makes it easier to satisfy the condition (\ref{equation1}).

%Length L and r Figure3 next two paragraphs
We want to understand the influence of $L$ and $r$ on our prediction algorithm in more detail. For this we determine the correlation between the original $x(t)$ time series  of Lorenz with predictions of the next 50 time units, while varying $L$ or $r$. Moreover we use these experiments to demonstrate that the location of the local section $\Sigma$ is of minor importance for the predictions. For each of our experiments we therefore use 10 random positions for $\Sigma$ and report in Fig.~\ref{figure3} the mean value as well as the 50\% confidence interval. For Fig.~\ref{figure3} (a) we fixed $w=2$, $L=150$ and vary the size of the partition, while in (b) we chose $w=10$, $r=3$ and vary $L$.   

\begin{figure}[bt]
\centering
\includegraphics[width=8cm]{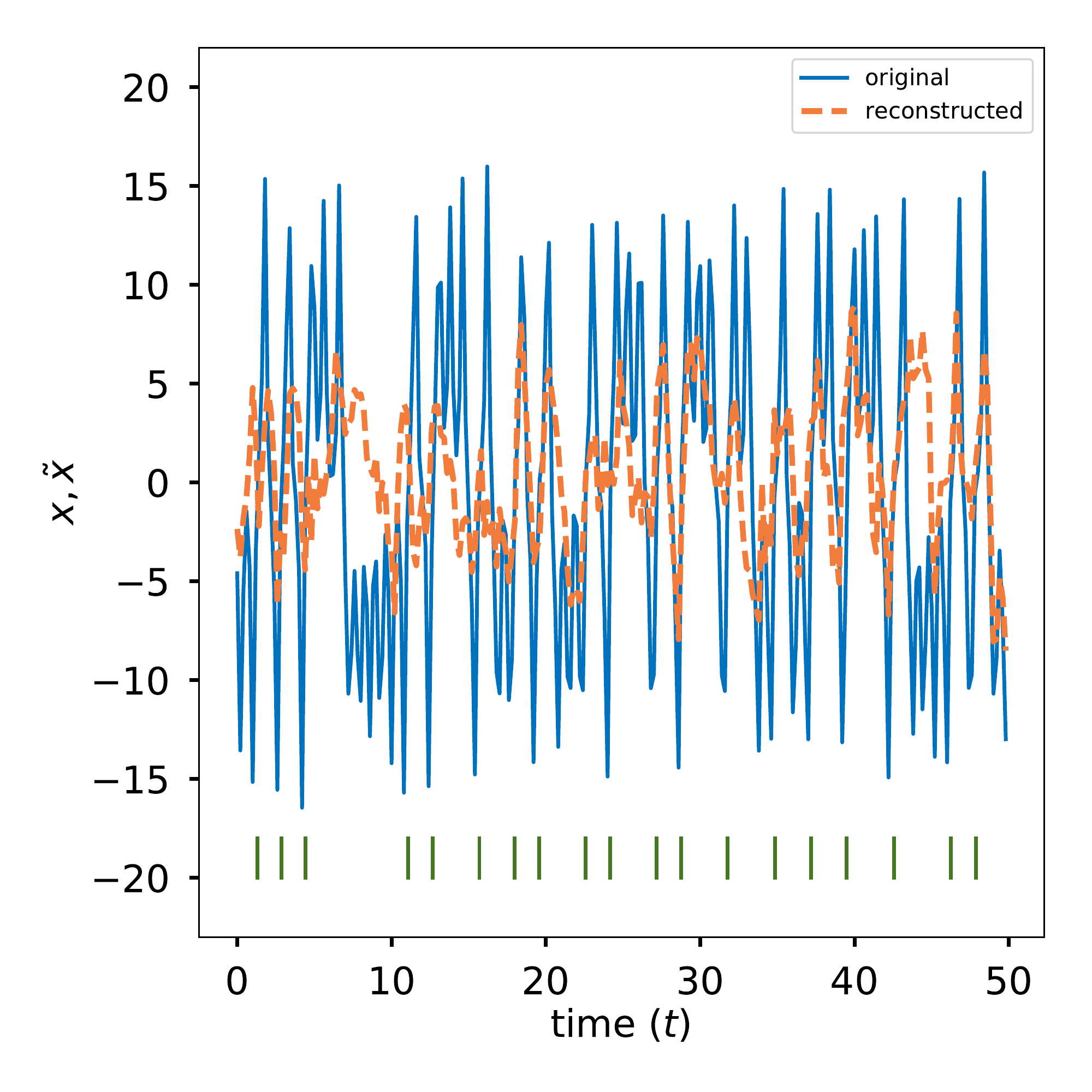}
%\caption{\label{figure5}\textcolor{black}{The local cross section at $x=0$ and $dx/dt > 0$ (panel a), its spike train and the reconstruction of the original signal (panel b). In panel b, the original signal is shown in the blue solid line and its reconstruction, the red dash-dotted line. Here we used $L=450$, $\Delta t = 0.2$ and $w = 30$. At the bottom we show the spike train and reconstruction for the local cross section at $x=0$ but $dx/dt<0$. The correlation coefficient between the original and reconstructed signals is 0.2273.}}
\caption{\label{figure5}\textcolor{black}{
A spike train generated by the local cross section at $x=0$ and $dx/dt > 0$, and the reconstruction of the original signal. The original signal is shown in the blue solid line and its reconstruction, the red dashed line. Here we used $L=19950$, $\Delta t = 0.2$ and $w = 30$. At the bottom we show the spike train and reconstruction for the local cross section at $x=0$ but $dx/dt<0$. The correlation coefficient between the original and reconstructed signals is 0.3926. (Because we have a fewer events, we needed the larger $L$ and $w$ to have the better results.)}}
\end{figure}

A sufficiently large size of $\Sigma$ enables us to cross-predict the original time series almost perfectly (Fig.~\ref{figure3}(a)), because for large $r$ we have a high recurrence rate and the condition of eq.~(\ref{equation1}) is likely to be met. When we reduce the size of the local cross section, the correlation coefficient between the original time series and the cross-predicted time series decreases. The lower recurrence rate makes it impossible to satisfy the condition eq.~(\ref{equation1}) \textcolor{black}{, namely $d>2m$,} for all times\textcolor{black}{, leading to the worse prediction}. But, even in such a case, a sufficiently large recording time $L$ \textcolor{black}{as well as a larger $w$} leads to a high correlation coefficient (Fig.~\ref{figure3}(b)). Thus, we expect for a database of a sufficiently long recording time that we can still reconstruct the original time series faithfully. Moreover the medians and 50\% confidence intervals in Fig.~\ref{figure3}(a) and (b) show that the positions of the local cross sections are of less importance for our prediction algorithm. \textcolor{black}{For example, reconstructions from local sections at $x = 0$ with $\dot{x} < 0$  or $\dot{x} > 0$ are not so different from each other as shown in Fig.~\ref{figure5}.}

%Some analysis
To further understand \textcolor{black}{these results in Figs.~\ref{figure2} and \ref{figure3}} and gain some insight about the importance of the different measurements (recorded history in the database and crossing times), we analysed the \textcolor{black}{predictability of flows from crossing times as an information-theoretic problem}. Our analysis shows that when a local cross section becomes small and therefore generated less frequent events, the time resolution of our data becomes the dominating factor. For our analysis we partition our spike train time series into $\alpha$ bins of equal size. Let $\beta$ be the number of events within the spike train. Moreover each bin can only contain one spike. Then, the number of signals $N$ we can send \textcolor{black}{by a spike train} is
\begin{equation}
N = \left(
\begin{array}{c}
\alpha\\
\beta
\end{array}
\right) = \frac{\alpha!}{\beta!(\alpha-\beta)!}.
\end{equation}
Assuming that all signals are equally likely, we can calculate the amount of information in the series as~\cite{cover2006} 
\begin{equation}
\log_2 N = \log_2 \frac{\alpha!}{\beta!(\alpha-\beta)!}.
\end{equation}
Applying Stirling's approximation, we get
\begin{equation}
\begin{array}{cl}
&\log_2 N\\
\approx & \alpha \{-\frac{\beta}{\alpha}\log_2 \frac{\beta}{\alpha}-\frac{\alpha-\beta}{\alpha}\log_2 \frac{\alpha-\beta}{\alpha}\}\\
= & \beta \log_2 \frac{\alpha}{\beta}+(\alpha-\beta)\log_2 \frac{\alpha}{\alpha-\beta}\\
\geq & \beta \log_2 \frac{\alpha}{\beta}.
\end{array}
\end{equation}
Given that $\alpha\gg \beta$, we can assume that $\alpha$ and $\beta$ are related by $\alpha = \beta 2^{\gamma}$. Therefore the RHS of the equation can be written as
\begin{equation}
\label{info}
\beta \log_2 \frac{\alpha}{\beta} = \beta \log_2 2^{\gamma} = \beta \gamma.
\end{equation}
%\todo[inline]{Perhaps one sentence that explains why $\beta$ is time resolution.}
Since $\beta$ gives the total number of spikes in the sequence and we only allow one spike per partition, the maximum $\beta$ is directly related to the time resolution of the sequence. We conclude that therefore the time resolution given by $\beta$ is directly proportional to the total amount of information contained in the time series. This has two consequences. For real world data the time resolution of the data is of upmost importance. Using our algorithm in real world applications would require instruments with high precision, but at least it is only the time precision and not also amplitude precision on top. The second consequence of eq.~(\ref{info}) relates to the efficiency of our method. Once the database is recorded, the data required to predict the future dynamics has a very small footprint. For example the database needed to predict the dynamics shown in Fig.~\ref{figure2}(b) has $40,000$ bytes and its crossing time series $\{t_i\}_{DB}$ has $152\pm40$ bytes \textcolor{black}{without compression in disk space (We used here the MATLAB command ``whos'' to evaluate the disk space we need for storing these variables.)}. The spike train recorded to estimate the state and consequently predict the dynamics was $0.38 \pm 0.10$ \% of their combined size. Having such a small memory demand makes this method attractive for real world applications.

\begin{figure}[tb]
\includegraphics[width=\linewidth]{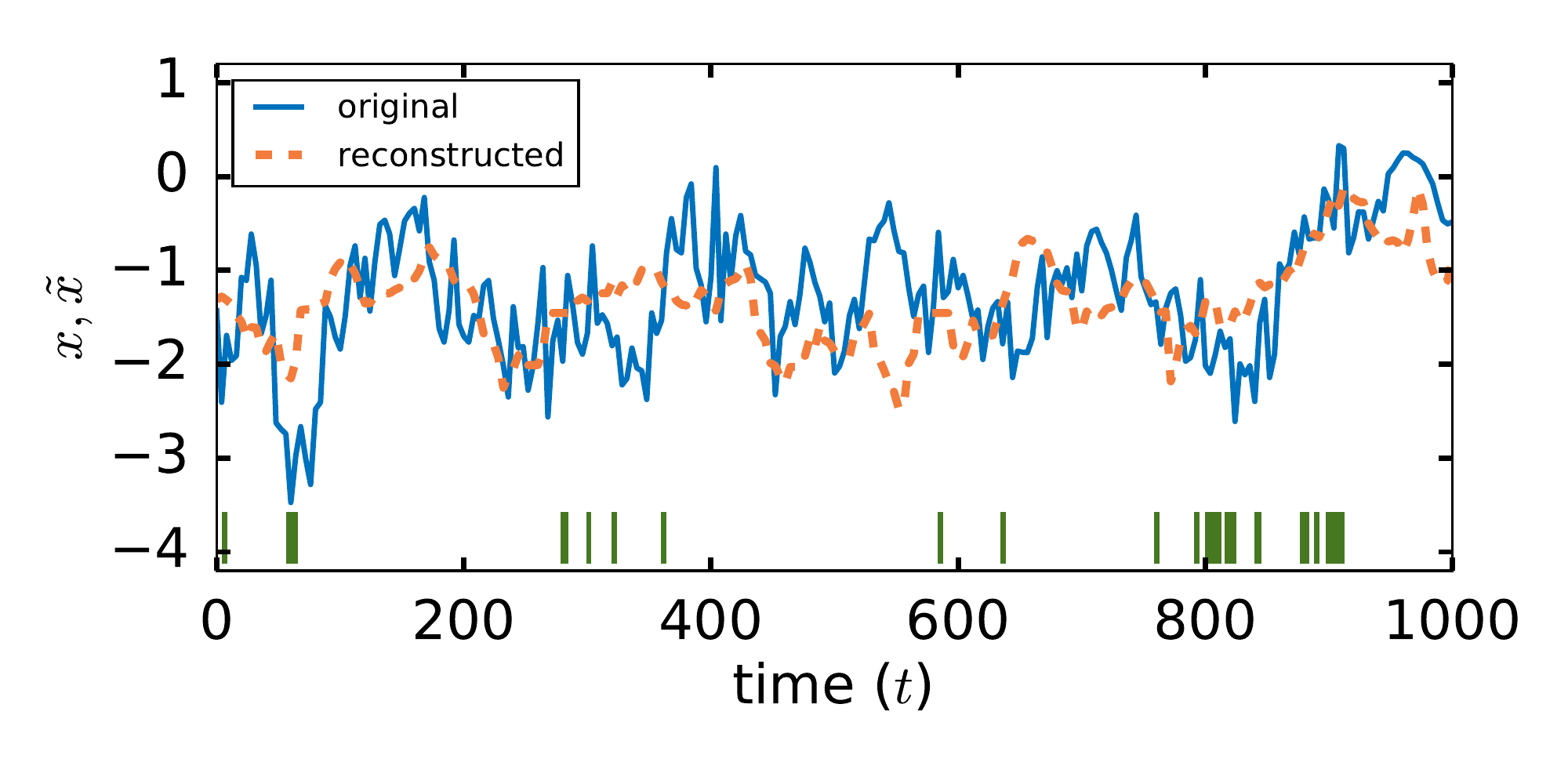}
\caption{\label{figure4}Example of the wind speed time series (solid line) together with our prediction (dash-dotted line). The parameters of our algorithms are the size of the time window $w=200\text{s}$, recording time $L=9,000\text{s}$, \textcolor{black}{$\Delta t=4\text{s}$} and the size of the local section $\Sigma=5\text{m}/\text{s}$. Our algorithm predicts the following 1,000 seconds.}
\end{figure}

%Here are dragons: the real world.
As one example for a real world application, we are use our algorithm to predict wind speed~\cite{hirata2008}. The original measurements were observed for 24 hours from around 2pm on 1 September 2005 at about 1m above from the ground level at the Institute of Industrial Science, The University of Tokyo, Tokyo, Japan using an ultrasonic anemometer.
%TIME(DATE) in WHERE using a MACHINE.
%\todo[inline]{A bit more details would be good, such as location, time (date), length of data, purpose, measurement equipment etc.}
The original measurements were made with 50Hz, but we first sub-sample the time series using every tenth point. This sub-sampling allows us to use a longer $L$ and therefore the database contains more of the long term changes in the dynamics. Since we do not have access to the full state space of this environmental system, we have to reconstruct the attractor of the dynamics by using delay embedding~\cite{takens1981,sauer1991}. Given that the system seems to have features of high-dimensional dynamics~\cite{hirata2015srep}, we have to use a high dimensional embedding. But the total recording time of the wind speeds is 10,000 seconds, and consequently we cannot use too high an embedding dimension, without making the recurrence rate too low for our prediction algorithm. In addition we only use the first 9,000 seconds as our database and use the remainder for comparison with our prediction. As a compromise and after several tests, we decided to use a 10 dimensional embedding space.  We find that our algorithm succeeded in capturing the large scale features of the dynamics, i.e., prolonged changes of the wind velocity are similar in the experimental data and the prediction (Fig.~\ref{figure4}). On the other hand our predictions fail to reproduce the finer details of the system's dynamics. We conjecture that a higher embedding dimension together with a longer time series would be able to overcome these limitations. Another way to improve the performance of the method could be to use marked spike trains, by assigning some additional value to each event. Several distances have been proposed for such marked point processes~\cite{schoenberg2008,suzuki2010,hino2015,iwayama2017}, which might increase the prediction performance for such high dimensional dynamics. 

%{\color{red}
Applying the method of Ref.~\cite{kennel1997} to the above wind data shows the value greater than 99\% point of $2.36$, meaning that the wind data should be regarded as non-stationary. The applicability of our method for non-stationary data should be evaluated more closely in future research, although our results in Fig.~\ref{figure4} seem to show some promise in this line of research.
%}

%Generating partions and Sigma
Our numerical results and to some extent the results on wind speeds show that a local cross section $\Sigma$ has the generic property to reconstruct the underlying dynamics of a flow almost perfectly. This is similar to generating partitions which have been extensively studied in maps~\cite{grassberger1985,davidchack2000,plumecoq2000a,plumecoq2000b,kennel2003,hirata2004}. In maps it is a non-trivial task to find a generating partition. Our results show, that in flows this is much easier and the position of $\Sigma$ does not matter much for the prediction quality (c.f. Fig.~\ref{figure3}). 

%We are different
While there are many studies that try to predict the next crossing time based on previous crossings (for example \cite{sauer1994}), the method presented here is different. Instead of just focussing on the next crossing, we are able, using cross-prediction, to estimate the complete future dynamics. If we regard the observations at a local cross section as a coincidence detector~\cite{konig1996,azouz2000} our results could explain how we can share the same experience even if we perceive a phenomenon in different ways, or in our case by different local cross sections. Hence, our results might also have some implications in the field of theoretical neuroscience.

%Recap and outlook
We presented a method for state estimation and prediction of flows, given a database of the past dynamics and a crossing time series. Identifying the current state is done by ranking the crossing time sequences of the database according to their distance from the current crossing time series. We showed that our method requires little memory \textcolor{black}{to store and send time series information via a spike train if the database is shared between sender and receiver}. Our method does not require us to store and send an entire time series \textcolor{black}{under this assumption}. Instead we simply record or send the times when a trajectory passes the local cross section. This advantage can be important for sensor networks~\cite{akyildiz2002}, where each device has to store and communicate lots of environmental information under severe energy constraints. Similarly, our method may be used to reconstruct missing data within observations, such as gaps in, e.g., satellites images partly covered by clouds~\cite{liu2017}, historical series of sunspot numbers~\cite{clette2014} or historical phenological data such as the start of the cherry blossom in Japan~\cite{aono2008}.

Y.~H.~would like to appreciate Prof. Kazuyuki Aihara for the discussions. The research of Y. H. is supported by Kozo Keikaku Engineering Inc. The research of D.~E.~has been supported by German-Israeli Foundation for Scientific Research and Development (GIF), GIF Grant No. I-1298-415.13/2015.

\end{document}